\documentclass[aps,prl,twocolumn,showpacs,superscriptaddress]{revtex4}

\usepackage{epsfig}
\usepackage{graphics}

\usepackage{natbib}

\begin{document}

\title{Fermi surface and quasiparticle excitations of Sr$_{2}$RhO$_{4}$}

\author{F. Baumberger}
\affiliation{Departments of Applied Physics, Physics, and Stanford Synchrotron Radiation Laboratory, Stanford University, Stanford, California 94305, USA}
\author{N.J.C. Ingle}
\altaffiliation{present address: Department of Physics and Astronomy, University of British Columbia, Vancouver, BC V6T 1Z4, Canada}
\author{W. Meevasana}
\author{K.M. Shen}
\altaffiliation{present address: Department of Physics and Astronomy, University of British Columbia, Vancouver, BC V6T 1Z4, Canada}
\author{D.H. Lu}
\affiliation{Departments of Applied Physics, Physics, and Stanford Synchrotron Radiation Laboratory, Stanford University, Stanford, California 94305, USA}
\author{R.S. Perry}
\affiliation{School of Physics and Astronomy, University of St. Andrews, St. Andrews, Fife KY16 9SS, Scotland}
\author{A.P. Mackenzie}
\affiliation{School of Physics and Astronomy, University of St. Andrews, St. Andrews, Fife KY16 9SS, Scotland}
\author{Z. Hussain}
\affiliation{Advanced Light Source, Lawrence Berkeley National Laboratory, Berkeley, California 94720, USA}
\author{D.J. Singh}
\affiliation{Condensed Matter Sciences Division, Oak Ridge National Laboratory, Oak Ridge, TN 37831-6032, USA}
\author{Z.-X. Shen}
\affiliation{Departments of Applied Physics, Physics, and Stanford Synchrotron Radiation Laboratory, Stanford University, Stanford, California 94305, USA}


\date{\today}

\begin{abstract}
The electronic structure of the layered 4$d$ transition metal oxide Sr$_{2}$RhO$_{4}$ is investigated by angle resolved photoemission. We find well-defined quasiparticle excitations with a highly anisotropic dispersion, suggesting a quasi-two-dimensional Fermi liquid like ground state. Markedly different from the isostructural Sr$_{2}$RuO$_{4}$, only two bands with dominant Rh 4$d_{xz,zy}$ character contribute to the Fermi surface. A quantitative analysis of the photoemission quasiparticle band structure is in excellent agreement with bulk data. In contrast, it is found that state-of-the-art density functional calculations in the local density approximation differ significantly from the experimental findings. 
\end{abstract}


\maketitle
The Fermi surface (FS) topology and quasiparticle dynamics determine most material properties. 
Low--dimensional and correlated materials, which are currently of key interest for their exotic properties, are particularly sensitive to fine details of the fermiology. 
This is evident for classical charge density wave systems, but might hold as well for superconductivity or quantum critical phenomena. 
For instance calculations for a large number of $p$-type cuprates, demonstrated a correlation of T$_{c}$ with the shape of the most bonding band \cite{pav01}. More recently, quantum criticality in Sr$_{3}$Ru$_{2}$O$_{7}$ has been related to a symmetry breaking spin--dependent FS distortion \cite{gri04ea}.\\
In principle, angle resolved photoemission (ARPES) is ideally suited to map the size and shape of the FS \cite{aeb94,dam03}. However, ARPES is a highly surface sensitive technique and its precision is usually lower than that of classical FS probes based on the de Haas--van Alphen (dHvA) or related effects. Consequently, the impact of ARPES has been largest in materials where dHvA oscillations are not observable \cite{dam03}. 
Density functional (DFT) calculations in the local density approximation (LDA) have evolved as a powerful alternative to experimental electronic structure probes.
However, the Kohn-Sham eigenvalues of DFT have no clear physical meaning even at the Fermi surface and cannot be rigorously identified with single particle excitation energies \cite{jon89}. Nonetheless, the LDA has been found to be highly successful even in the description of fairly strongly correlated materials like doped cuprates.
Although the agreement of LDA calculations with extensive ARPES data on cuprates is compelling, it has rarely been confirmed by bulk electronic structure probes, and truly quantitative comparisons are not ready yet.
Sr$_{2}$RuO$_{4}$ is to date the only example of a correlated oxide where LDA \cite{ogu95,sin95} dHvA \cite{mac96,ber00} and ARPES  \cite{dam00ea} were found to be in good quantitative agreement.
This is far from trivial in a multi--band system, since
correlations, not fully described within the LDA, can transfer spectral weight between in equivalent orbitals, thus enlarging certain FS pockets at the expense of others \cite{lie00,ish05}.
Na$_{x}$CoO$_{2}$ is a prominent recent example of a multi--band system where qualitative differences between ARPES data \cite{has04ea,yan05ea} and LDA calculations \cite{joh04} have been observed and related to strong, orbital dependent correlations. However, the complexity of the material has so far prevented the derivation of a consistent picture \cite{ish05,zho05}.

In this paper, we present a quantitative electronic structure study of Sr$_{2}$RhO$_{4}$ by means of ARPES and band calculations within the LDA. It is shown that ARPES provides bulk representative spectra with a FS that agrees with dHvA data by Perry $\textit{et al.}$ \cite{per06ea}. Although Sr$_{2}$RhO$_{4}$ exhibits Fermi--liquid properties over an extended energy range, we find that its FS is not reproduced quantitatively within the LDA.

\begin{figure}[tb]
\includegraphics[width=0.38\textwidth]{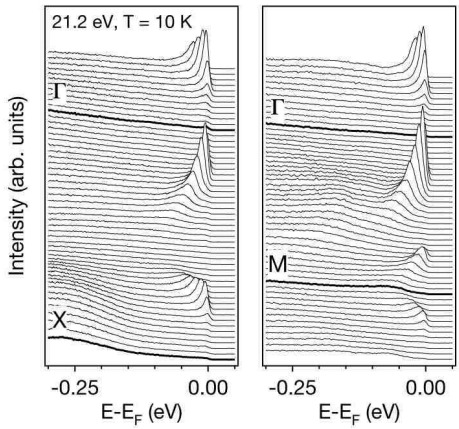}
\caption{\label{f1} ARPES spectra along $\Gamma$X and $\Gamma$M of the orthorhombic Brillouin zone. Spectra taken at $\Gamma$, X and M are highlighted. For a definition of the symmetry points, see Fig. 2.}
\end{figure} 
%

Sr$_{2}$RhO$_{4}$ has a tetragonal crystal structure ($a=5.436$~\AA, $c=25.75$~\AA) with a reduced $I4_{1}/acd$ symmetry ("orthorhombicity") due to a 11$^{\circ}$ rotation of the RhO$_{6}$ octahedra around the c-axis \cite{shi92, vog96}. 
High purity single crystals with residual resistivities $< 7\mu\Omega$cm have been grown by a floating zone technique \cite{per06ea} and have been cleaved $\textit{in situ}$ along the $ab$--plane at $T=10$~K. Photoemission experiments were performed with a monochromatized He--discharge lamp (Gammadata VUV5000) and a Scienta SES2002 analyzer. The energy and angular resolutions for all measurements were better than 7.5~meV / $0.35^{\circ}$ [full width at half maximum (FWHM)]. 
All data were taken at $T = 10$~K and a pressure $< 4\times10^{-11}$~torr.
LDA calculations were done using the local orbital extension of the general potential linearized augmented planewave method \cite{singh} with well converged basis sets ($\approx3800$ basis functions) and zone samplings.

\begin{figure}[bt]
\includegraphics[width=0.49\textwidth]{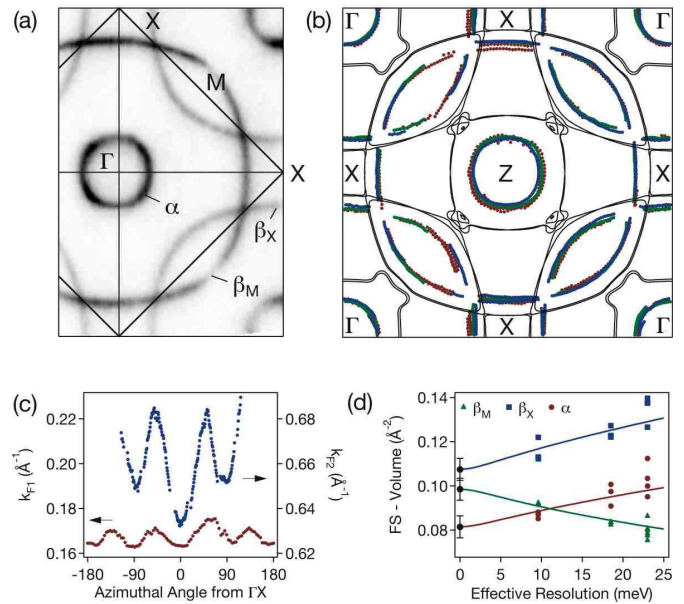}
\caption{\label{f2} (a,b) Experimental (a) and theoretical (b) LDA--FS calculated for a fully relaxed crystal structure of Sr$_{2}$RhO$_{4}$. X denotes the surface projection of the X--point of the tetragonal unit cell, M the projection of the midpoint between $\Gamma$ and Z. Experimental FS contours have been extracted in areas where peaks are well separated from data-sets taken with $h\nu=21.2$~eV (red and blue dots) and 40.8~eV (green) and are overlaid on the theoretical FS. (c) Fermi wave vectors of the two fundamental bands in the unfolded tetragonal BZ, showing a 4--fold modulation. (d) Resolution dependence of the apparent FS--volume. Symbols give the apparent volumes of the three pockets, measured with different energy resolutions, lines are obtained from an analysis of simulated spectral functions that have been convoluted with varying resolution functions. Extrapolated FS volumes are shown as black circles with estimated error bars.}
\end{figure} 
\begin{figure}[tb]
\includegraphics[width=0.45\textwidth]{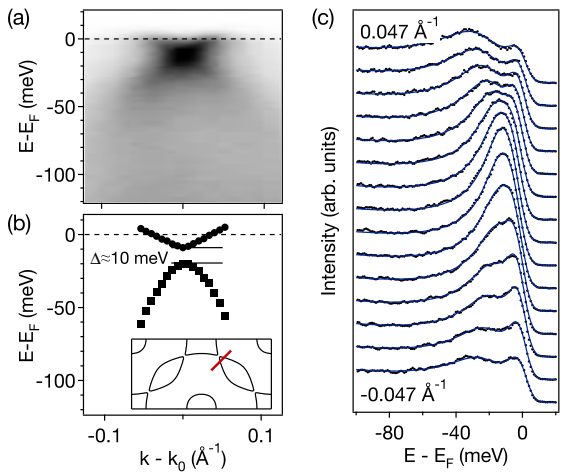}
\caption{\label{f1} Band dispersion near the orhorhombic zone boundary measured along the red line depicted in the Brillouin zone inset to panel (b). (a) and (c) show the raw data as image plot and stack of EDCs respectively. A highly restricted fit with only two free parameters per spectrum is added as thin blue lines in (c). (b) shows the fitted band dispersion, with a non--crossing gap of approximately 10~meV, due to the orthorhombicity.}
\end{figure} 
Representative spectra along $\Gamma$X and $\Gamma$M of the orthorhombic Brillouin zone (BZ) are shown in Fig. 1. The data clearly show the spectroscopic hall marks of a Landau Fermi liquid: well defined, dispersive quasiparticle bands with peaks that sharpen up progressively as the they approach the Fermi level ($\rm{E_{F}}$), reflecting the diminishing phase space for electron--electron scattering \cite{cla92}. 
Fig. 2(a) shows the experimental FS map, obtained from $\approx 2\cdot10^{4}$ high--resolution spectra, taken on a uniform k--space grid and integrated over the energy window $\rm{E_{F}}\pm3$~meV. Note that the resulting map has not been symmetrized. Only the measured momentum--space region is shown.
Similar data sets were measured on several samples with photon energies of 21.2~eV (He I$\alpha$) and 40.8~eV (He II$\alpha$) and showed good reproducibility and no excitation--energy dependence, consistent with a highly two--dimensional (2D) electronic structure.	
Two, nearly isotropic bands, which are both centered at the origin and back--folded by the orthorhombicity are observed, a large electron like band with an average Fermi wave vector $k_{F2}$ of $\approx$~0.66~\AA$^{-1}$ and a smaller hole--pocket with $k_{F1}\approx0.17$~\AA$^{-1}$. A quantitative determination of the unfolded Fermi wave vectors (Fig. 2(c)) shows a slight 4--fold anisotropy, analogous to the shape of the $\alpha/ \beta$ sheets in Sr$_{2}$RuO$_{4}$, and indicative of a dominant $d_{xz,zy}$ character of the FS.
The LDA calculation (Fig. 2(b)) confirms this experimental assignment
and shows that the $d_{xy}$ level is pushed below $E_{F}$ by level mixing and repulsion between the $e_{g}$ $d_{x{^2}-y{^2}}$ and the $t_{2g}$ $d_{xy}$ orbitals, as discussed in Ref. \cite{kim06ea}.
At the orthorhombic zone boundary, a small non--crossing gap opens. 
This is investigated in more detail in Fig. 3 showing the band dispersion across the zone boundary.
The gap is not directly resolved in the raw data, but the flattened peak shape of the energy distribution curves (EDCs) shown in Fig. 3 (c) hints at the presence of a non-crossing gap slightly smaller than the line width of about 20~meV at the energy where the two branches intersect. For a more quantitative analysis, we fit the data with the spectral function for a 2D Fermi liquid superimposed on a smooth background. To increase the reliablility of this analysis, we choose to fit all EDCs simultaneously with a common self--energy and a momentum independent intensity and convolved the fit--function in energy and momentum with the independently determined respective resolutions. The resulting band positions are shown in Fig. 3(b) to display a gap of $\approx10$~meV. This is significantly larger than the Landau level splitting even in high magnetic fields. 
Thus, in a 2D approximation, the FS contains three closed contours, a central hole pocket ($\alpha$), the lens--shaped electron pockets at M ($\beta_{M}$), and the square--shaped hole pockets at X ($\beta_{X}$).

\begin{table}[htb]
\caption{\label{Rh-table}Summary of the ARPES FS parameters for Sr$_{2}$RhO$_{4}$. The errors given throughout the paper are estimated from the statistical accuracy of the analysis and the reproducibility of the experiments. A systematic error of the same order due to surface structural relaxations cannot be excluded. The frequency range of dHvA oscillations (from Ref. \cite{per06ea}) and $k_z$--averaged LDA volumes are added for comparison.
}
\begin{ruledtabular}
\begin{tabular}{l c c c}
 & $\alpha$ & $\beta_{M}$ & $\beta_{X}$ \\
 \hline
FS--volume $A$ (\% BZ) & 6.1(4) & 7.4(4) & 8.1(5) \\
occupation $n$ ($e^{-}$) & 3.878(8) & 0.296(16) & 1.838(10) \\
Fermi velocity $v_{F}$ (eV\AA) & 0.41(4) & 0.61(6) & 0.55(6) \\
cyclotron mass $m^{*}$ ($m_{e}$) & 3.0(3) & 2.2(2) & 2.6(3) \\
\hline
dHvA FS--vol. $A$ (\% BZ) & &6.6 -- 9.2& \\
LDA FS--vol. $A$ (\% BZ) & $\approx 24$ & $\approx 14$ & $\approx 4$ \\
\end{tabular}
\end{ruledtabular}
\end{table}

We have determined the volume $A$ of these pockets from extensive fits to multiple data sets to be 6.1(4), 7.4(4), 8.1(5)\% BZ for $A_{\alpha}$, $A_{\beta_{M}}$ and $A_{\beta_{X}}$, respectively, as 
summarized in Tab. \ref{Rh-table}. Each of these values represents the average of 8--10 equivalent pockets, measured on different samples and with different photon energies and energy resolutions, and contains a correction for the systematic shifts in the zero frequency MDC peak positions (see Fig. 2(d)) caused by the finite energy resolution of the experiment \footnote{Details of the data-analysis will be published elsewhere.}.  
The such derived values are in excellent agreement with the frequency range of 6.6 to 9.2\% BZ, in which dHvA oscillations have been observed very recently \cite{per06ea}.
In order to estimate the $total$ Luttinger volume, we assume two--dimensionality. After backfolding the fundamental bands to the orthorhombic Brillouin zone, which contains two Rh atoms per plane, we count $n_{\alpha}=2+2-2A_{\alpha}=3.878(8) e^{-}$ electrons for the hole--pocket at $\Gamma$, $4A_{\beta_{M}}=0.296(16) e^{-}$ in the lens--shaped electron pocket, and $2-2A_{\beta_{X}}=1.838(10) e^{-}$ for the X--point hole pocket. The three pockets thus contain 3.006(10) electrons per Rh, consistent with a stoichiometric material and a fully occupied $d_{xy}$--band.

Carrier masses have been determined using Fermi velocities evaluated in typically 100 $I(\epsilon,k)$ intensity distributions along $k$--space lines normal to the FS contour of each pocket.
The averaged $v_{F}$ values are given in Tab. \ref{Rh-table}. The cyclotron masses $m^{*}=hk_{F}/v_{F}$ are then calculated for average Fermi wave vectors $k_{F}=\sqrt{A/\pi}$ to be $m^{*}_{\alpha}=3.0(3)$, $m^{*}_{\beta_{M}}=2.2(2)$, and $m^{*}_{\beta_{X}}=2.6(3)$.
Again assuming two dimensionality, the specific heat is given by $\gamma=(\pi N_{A}k_{B}^{2}a_{o}^{2})/(3\hbar^{2})\sum{m^{*}}$, where $k_{B}$ is the Boltzmann constant, $N_{A}$ Avogadro's number, and $a_{o}$ the in--plane lattice constant. Accounting for the twofold degeneracy of the $\beta_{M}$--pocket, we find $\sum{m^{*}}=10.0(5) m_{e}$, and $\gamma=14.5(7)$~mJ/molK$^{2}$, in fair agreement with the preliminary experimental report of 17.7(7)~mJ/molK$^{2}$ \cite{per06ea}.

%
\begin{figure}[tb]
\includegraphics[width=0.47\textwidth]{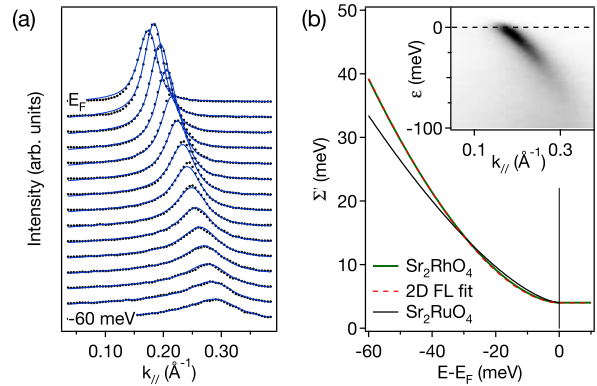}
\caption{\label{f1} Spectral function analysis of the $\alpha$ pocket along $\Gamma$M. (a) experimental MDCs with the result of a 2D fit (thin blue lines). (b) Comparison of the self--energies for Sr$_{2}$RhO$_{4}$ and Sr$_{2}$RuO$_{4}$. The functional form derived for the latter in Ref. \cite{ing05} has been offset by 3~meV to match the slightly higher impurity scattering in Sr$_{2}$RhO$_{4}$. A fit of the analytical form for a 2D FL to the empirical $\Sigma'$ derived for Sr$_{2}$RhO$_{4}$ is shown as red dashed line. The inset shows a gray--scale plot of the measured spectral function.}
\end{figure} 

The QP self--energy has been derived from a detailed line--shape analysis summarized in Fig. 4. We first note that there exists no simple relation between the width of a single MDC (or EDC) and the self--energy for sharp and non--linear QP--bands as observed in Sr$_{2}$RhO$_{4}$, even for high--resolution data as those presented here. Therefore, we chose to perform 
2D fits with a parametrized self--energy $\Sigma'=\Sigma_{imp}+\beta\omega^{\eta}$, as well as simultaneous 1D fits of all MDCs. Both of these methods are fully self--consistent and allow one to include a convolution with both, the energy and momentum resolution functions. 
The results of a typical 2D fit are shown in Fig. 4(c) and demonstrate that widths, asymmetries and intensities of the MDCs are well reproduced with a minimal parameter set. The imaginary part of the self--energy deduced in this way is $\Sigma'=0.004~eV+4.7\omega^{1.74}$, in close agreement with the analytical form $\Sigma'(\omega)=\beta\omega^2[1+0.53 \ln(\omega/E_{F})]$ for a 2D Fermi liquid with realistic parameter values $\beta=4.8$ and $E_{F}=0.43$~eV (see Fig. 4(b)) \cite{hod71}. 
Moreover, it is nearly identical with the recent result for Sr$_{2}$RuO$_{4}$ \cite{ing05}, hinting at comparable many--body interactions in both materials, dominated by electron--electron interactions.
The strength of electron--phonon interactions cannot be determined reliably from the present data, since the coherent QP peaks can only be separated over a limited energy range barely larger than typical phonon frequencies.
However, the quality of the above fit with a smooth QP--dispersion up to $\omega=60$~meV and a form of $\Sigma'$ expected for electron--electron interactions only seems to
indicate a minor importance of other degrees of freedom.

The correct volume counting of the expected number of electrons, the agreement with dHvA data, and the sheer observation of single, sharp QP peaks show that the 2-3 topmost unit cells of Sr$_{2}$RhO$_{4}$ which are probed by ARPES have a uniform and basically converged bulk electronic structure (for an example of ARPES data from a reconstructed surface, see $\textit{e.g.}$ Ref. \cite{she01ea}). We therefore use the zero frequency line width of $\approx8$~meV (corresponding to $\Delta k_{F}\approx 0.015$\AA$^{-1}$, $\Delta A\approx 1.2$\% BZ) as an upper bound for possible systematic errors in the ARPES $\epsilon(k)$ values caused by surface structural relaxations. 
This uncertainty is far smaller than the difference between ARPES and LDA Fermi wave vectors. The presented results thus establish for the first time a quantitative discrepancy between the experimental quasiparticle FS and the DFT--LDA FS in a Fermi--liquid--like correlated material.
The two main discrepancies between calculation and experiment are the shape of the $\alpha$ pocket and the volume ratios between the three main pockets, with LDA finding values of $A_{\alpha}\approx 24$\% BZ, $A_{\beta_{M}}\approx 14$\% BZ, and $A_{\beta_{X}}\approx 4$\% BZ, clearly incompatible with both, ARPES and dHvA. 
The disagreement could perhaps be explained by a
deviation of the crystal structure from that
assumed in the calculations or by an O deficiency.
However, there is no experimental evidence for either.
The Luttinger volume of 3.006(10) indicates good stoichiometry, and
calculations performed with the experimental lattice structure from Ref. \cite{vog96}, artificially distorted structures, and the fully relaxed LDA crystal structure all revealed rather similar LDA--FS that differ significantly from the experiment. We also searched for symmetry reducing rotations of the RhO$_{6}$ octahedra, but found that only the 
reported $I4_{1}/acd$ symmetry \cite{shi92,vog96} without out of plane tilt--distortions is stable within LDA.

It is thus compelling to attribute the observed discrepancy between LDA and ARPES to many--body interactions, not fully described within the LDA. 
The functional form of $\Sigma'$ shown in Fig. 4. indicates that these interactions are dominated by electron--electron scattering,
although it cannot fully exclude a more complex interplay of correlations and electron--phonon coupling, as it is $\textit{e.g.}$ observed in the metallic 4$d$ compound Ca$_{3}$Ru$_{2}$O$_{7}$ \cite{bau06ea}. 
The structural distortions in Sr$_{2}$RhO$_{4}$ may well be crucial for the marked difference in experimental and LDA FS since they reduce the band width and Fermi velocity by nearly a factor of two, compared to Sr$_{2}$RuO$_{4}$. Consequently, for Sr$_{2}$RhO$_{4}$, a band dependence of the real part of the self--energy at $\omega=0$ of $\approx 120$~meV as it was calculated by Liebsch and Lichtenstein for Sr$_{2}$RuO$_{4}$ \cite{lie00} would be sufficient to explain the discrepancy between ARPES and LDA FS reported here.

\begin{acknowledgments}
We thank B.J. Kim and C.Y. Kim for discussion and for provision of Ref. \cite{kim06ea} prior to publication. 
This work has been supported by the ONR grant N00014-01-1-0048. Additional support from SSRL is provided by the DOE's office of Basic Energy Science, Division of Material Science with Contract DE-FG03-OIER45929-A001.
\end{acknowledgments}


\end{document}